\documentclass[12pt]{article}
\usepackage[space]{scicite}
\usepackage{times}
\usepackage{graphicx}
\usepackage{url}

\newenvironment{sciabstract}{%
\begin{quote} \bf}
{\end{quote}}

\newcounter{lastnote}

\title{\textbf{Electronic zero-point fluctuation forces inside circuit components}}
\author{Ephraim Shahmoon$^1$ and Ulf Leonhardt$^2$\\
\normalsize{$^1$Department of Physics, Harvard University, Cambridge MA 02138, USA}\\
\normalsize{$^2$Department of Physics of Complex Systems,}\\
\normalsize{Weizmann Institute of Science, Rehovot 761001, Israel}}
\date{\today}

\begin{document}
\maketitle

\begin{sciabstract}
One of the most intriguing manifestations of quantum zero-point fluctuations are the van der Waals and Casimir forces, often associated with vacuum fluctuations of the electromagnetic field. Here we study generalized fluctuation potentials acting on internal degrees of freedom of components in electrical circuits. These electronic Casimir-like potentials are induced by the zero-point current fluctuations of any general conductive circuit. For realistic examples of an electromechanical capacitor and a superconducting qubit, our results reveal the possibility of tunable forces between the capacitor plates, or the level shifts of the qubit, respectively. Our analysis suggests an alternative route towards the exploration of Casimir-like fluctuation potentials, namely, by characterizing and measuring them as a function of parameters of the environment.
Such tunable potentials may be useful for future nanoelectromechanical and quantum technologies.
\end{sciabstract}

\section*{Introduction}

Understanding the role of quantum phenomena in electrical circuits has opened numerous possibilities in quantum information and optics \cite{SCH,GRV}. In this article we show how conceptually simple systems, such as linear electrical circuits, provide a new direction in the study and application of quantum fluctuation phenomena, in analogy to the van der Waals (vdW) and Casimir forces \cite{LON,CP,CAS}.
A key element of the ensuing discussion is the generalization of the familiar Casimir force in three aspects; namely, the source of the quantum fluctuations, the physical effect they entail and, most importantly, how the effect is measured and manipulated.

The Casimir and vdW forces are typically associated with quantum zero-point fluctuations of the electromagnetic field modes. A physical system coupled to these modes then acquires additional potential energy, which for a composite system, leads to a force between its components, {\it e.g.}\ a pair of particles or surfaces in the case of vdW or Casimir, respectively. Nevertheless, such a quantum-fluctuation-driven potential may exist for any system that can exchange energy with a quantum reservoir, following the fluctuation-dissipation theorem \cite{LIF,Ginzburg,MIL,FoQV,QEDFDT}. In the case of electrical circuits considered here, we employ this principle within the framework of circuit theory, by considering the quantum, zero-temperature noise emerging from resistive circuit elements, in analogy to the Johnson-Nyquist noise at finite-temperatures \cite{NYQ,BIM}.

The electromagnetic vacuum causes forces between polarizable objects \cite{CP,CAS,LAM,MOH,CAP,DEC,CAPr,PARA,FoQV,REV} and generalized forces such as torques \cite{PARS,MUNt,REYt,RAUL} and lateral forces \cite{KAR,MOHl,REYl,KIM}, or energy level shifts in atoms \cite{MIL,SCU}. In analogy, we consider here forces/torques acting on internal mechanical degrees of freedom of capacitive elements, such as between capacitor plates, and energy shifts in superconducting qubits; all of which driven by the electronic zero-point noise of the circuit (Fig.~1). We show that a macroscopic circuit-theory approach can be used to predict such zero-point potentials and forces. When divergences in these potentials occur, they are renormalized by subtracting the zero-point potential of a reference circuit; therefore, our approach often predicts the \emph{relative} forces, due to variations in circuit parameters. This approach is thus complementary to more microscopic formulations, sensitive to the physical details and material properties of circuit components \cite{WOODSr}, such as the recent work on the effect of charge fluctuations in capacitors made of graphene \cite{Abinitio}. One advantage of the macroscopic approach presented here is its capability to predict observable forces in realistic systems using simple lumped-element modeling. For example, considering the macroscopic experimental parameters of a recently-realized quantum electromechanical plate capacitor \cite{TEU,LEHN}, we find tunable relative forces between its plates (on top of the familiar Casimir force), which are within its current measurement sensitivities. We demonstrate that both the sign and spatial dependence of the relative force can be controlled by simply tuning the circuit parameters. This suggests that the manipulation and understanding of such electronic zero-point forces may become relevant for current quantum electromechanical devices, which are expected to play a role in hybrid quantum technologies \cite{CLE,CHU,ROU}.

An important opportunity opened by the exploration of zero-point fluctuation phenomena in circuits is the possibility to measure them as a function of various circuit parameters. Casimir-like forces are typically characterized and measured as a function of distance between the interacting objects \cite{FoQV}, whereas their surrounding electromagnetic environment remains unchanged between measurements. In contrast, here both the surrounding environment and zero-point fluctuations sources, that are realized by the reactive and resistive components of the circuit, respectively, can be varied between different measurements. Then, the potential ({\it e.g.}\ force) acting on an internal coordinate of a component embedded in the circuit, can be measured as a function of the parameters of the environment, while keeping the internal degrees of the component, {\it e.g.}\ the separation between capacitor plates, fixed. This new possibility may have several advantages as discussed below.

\section*{Results}

\subsection*{Electronic zero-point potential}

Consider first the source of quantum fluctuations in circuits. The electric conductors that comprise electronic circuits absorb the energy of electromagnetic fields. Hence, following the fluctuation-dissipation theorem, their electronic degrees of freedom must fluctuate. Within circuit theory, this can be accounted for by accompanying any resistance $R$ in the circuit by a shunt fluctuating current source, $I_N(t)$, with a zero mean and a spectrum (Fig.~1a)
\begin{equation}
S_{I_N}(\omega)=\int_{-\infty}^{\infty}dt\, e^{i\omega t}\langle I_N(t) I_N(0)\rangle=\frac{2\hbar\omega}{R(\omega)}\frac{1}{1-e^{-\hbar \omega/T}},
\label{IN}
\end{equation}
$T$ being the temperature. Such a treatment of quantum noise in circuits is equivalent to the explicit quantization of the circuit \cite{DEV}, as was recently shown also in the context of Casimir forces \cite{QEDFDT}. For large $T$ compared with $\hbar \omega$, the thermal noise is the well-known Johnson-Nyquist noise of resistors; here we focus on fluctuations in circuits at zero temperature and the forces they cause.

\begin{figure}
\begin{center}
\includegraphics[width=\columnwidth]{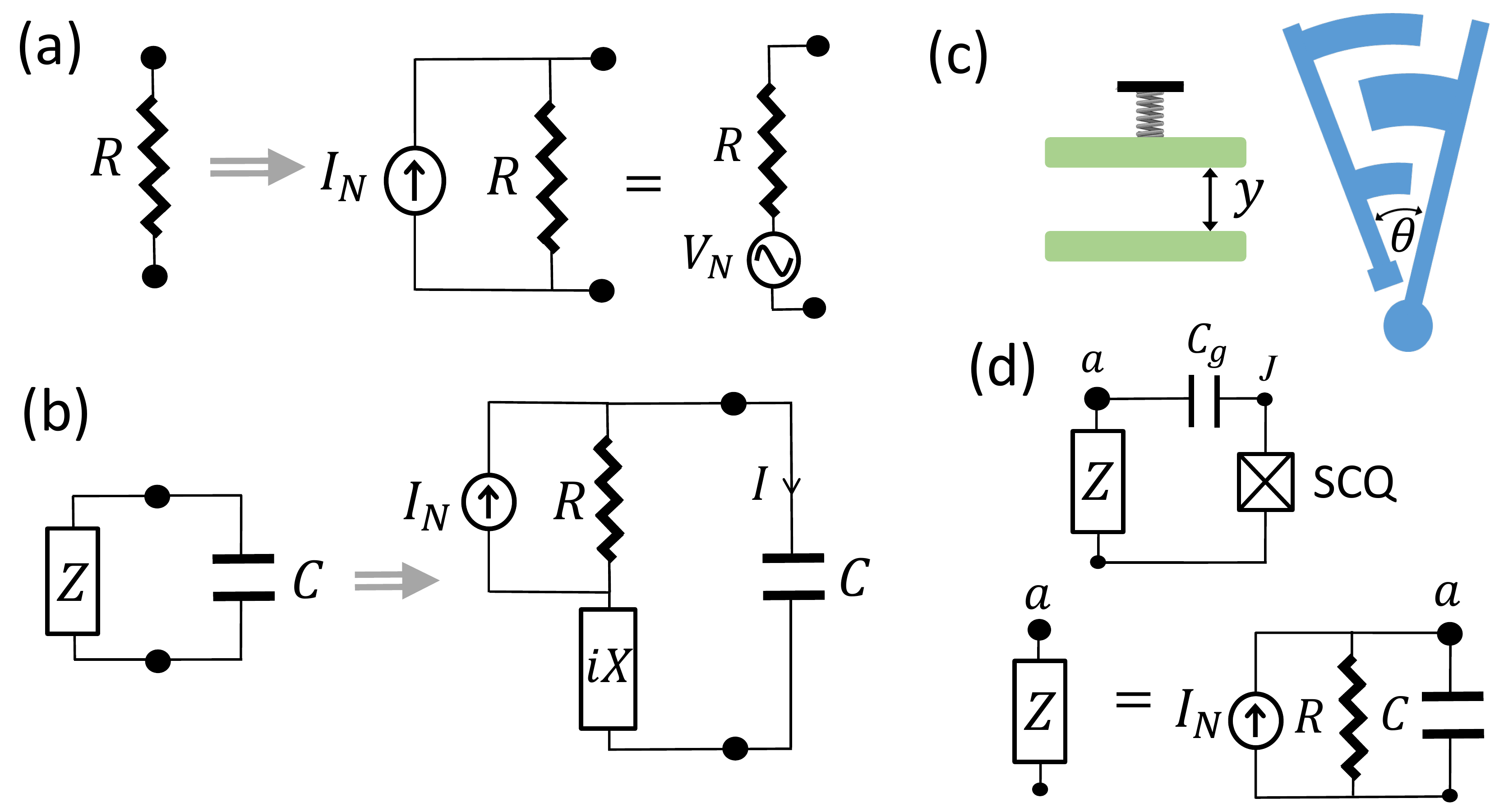}
\caption{\small{
Generalized potentials induced by electronic quantum fluctuations. (a) Quantum noise source: resistive circuit elements are modeled by a resistor $R$ in parallel with a current noise source $I_N$ with the spectrum of Eq.~(\ref{IN}) (or equivalently, in series with a voltage $V_N=R I_N$).  (b) Capacitor embedded in a general passive circuit represented by its impedance $Z=R+iX$. The zero-point-induced potential built on the capacitor, Eq. (\ref{U}), gives rise to a generalized force $f=-\partial U/\partial\xi$ on its internal degree of freedom $\xi$. (c) Examples of electromechanical capacitors: for the parallel-plate capacitor with separation $\xi=y$ \cite{TEU,LEHN}, $f$ is a force normal to the plates, whereas for the variable capacitor with rotation angle $\xi=\theta$ \cite{ROT}, $f$ is a torque.  (d) Superconducting qubit (SCQ)  capacitively coupled to a circuit $Z$ (bottom: specific example of an $RC$ circuit). The zero-point fluctuations from $Z$ induce shifts in the energy levels of the SCQ in analogy to the Lamb shift.
 }} \label{fig1}
\end{center}
\end{figure}

After having established the noise spectrum of the quantum fluctuations in a circuit with resistance (Fig.~\ref{fig1}a), consider now the resulting zero-point energy in a capacitor connected to the circuit (Fig.~\ref{fig1}b). The time-averaged electromagnetic energy stored inside the capacitor $C$ is given by
\begin{equation}
U=\frac{C}{2}\int_{-\infty}^{\infty}\frac{d\omega}{2\pi}S_V(\omega), \quad S_V(\omega)=\frac{\langle V(\omega)V(-\omega)\rangle}{T_{e}},
\label{Uc}
\end{equation}
where $V(\omega)=\int_{-\infty}^{\infty}dt\, e^{i\omega t}V(t)$ is the Fourier transform of the voltage $V(t)$ across the capacitor, $S_V(\omega)$ is its fluctuation spectrum and $T_e\rightarrow \int_{-\infty}^{\infty} dt=2\pi\delta(\omega=0)$ is the duration of the experiment. Therefore, the potential energy $U$ is determined by the spectrum of the voltage $V(\omega)$, that depends on the circuit in which the capacitor is embedded. In the absence of any external sources, a general linear circuit can always be characterized by its total impedance $Z(\omega)=R(\omega)+iX(\omega)$, $R$ being the resistive part and $X$ the reactive part (both real) with $Z(-\omega)=Z^{\ast}(\omega)$. For taking into account the quantum noise, we replace the resistance by the circuit of Fig.~1a.
The voltage across the capacitor is equal to the voltage over the left side of the circuit (Fig.~1b)
\begin{equation}
V=i\frac{I}{\omega C} = R\,(I_N-I) + iX I,
\label{voltage}
\end{equation}
where $I$ is the current charging the capacitor. We solve for $I$ in terms of $I_N$ and express $V(\omega)V(-\omega)$ of Eq.~(\ref{voltage}) in terms of the spectrum, Eq.~(\ref{IN}) for $T=0$. From Eq.~(\ref{Uc}) we then obtain
\begin{equation}
U =\hbar \int_0^{\infty}\frac{d\omega}{2\pi}\,\frac{R(\omega) C \omega}{[1+X(\omega)C\omega]^2+[R(\omega)C\omega]^2},
\label{U}
\end{equation}
which is the electronic zero-point fluctuation-induced energy stored in the capacitor.

It should be emphasized that the potential, Eq. (\ref{U}), was obtained using only elementary circuit theory with the addition of quantum noise, Eq. (\ref{IN}). Other potentials, such as the standard zero-point Casimir force between the capacitor plates, may exist in addition. From a quantum-electrodynamic point of view, the potential of Eq. (\ref{U}) is the contribution to the total zero-point potential, that is mediated and driven solely by quantum fluctuations of the fundamental transverse-electromagnetic mode guided by the circuit wires \cite{QEDFDT}. The standard Casimir force can then be attributed to quantum fluctuations of the rest of the electromagnetic modes, which are not accounted for by circuit theory and Eq. (\ref{U}) [see Supplementary Materials (SM)].

\subsection*{Renormalization and the relative force}

It turns out that in most cases the zero-point energy $U$ needs to be renormalized in order to extract the physically meaningful force: one sees in Eq.~(\ref{U}) that $U$ diverges if neither $X$ nor $R$ grow sufficiently with frequency. Note that the lumped-element description used in circuit theory is based on a quasistatic approximation valid for $\omega\ll c/l$, $l$ being a typical element size. So this possible divergence implies that circuit theory cannot account well for the total zero-point energy.

For the complete renormalization of the potential $U$, microscopic details, not accounted for by lumped-circuit theory, should be considered. These may include the physical details of the capacitor, such as in the excellent {\it ab-initio} paper \cite{Abinitio}, or the finite lengths and configuration of the connecting wires (see below). However, we show now that one does not have to give up the generality of the above macroscopic description in order to obtain an observable, finite result.
In fact, circuit theory {\it can} account for the \emph{relative} zero-point energy due to variations in circuit parameters, as long as the frequency range over which $X$ and $R$ differ is smaller than $c/l$. To give an example, consider a capacitor $C_0$ and a resistor $R$ in series (both independent of frequency) coupled to $C$ (case I in Fig.~2a). Inserting the reactance of the capacitor, $X=1/(\omega C_0)$, in Eq.~(\ref{U}) we see that the integral diverges logarithmically. However, if we take the difference between $U$ and the corresponding energy for $X=0$ --- the zero-point energy $U_{RC}$ of a pure $RC$ circuit (where $C_0$ is shorted) --- we obtain the difference between two logarithms, which gives the finite, exact result:
\begin{equation}
U_{\mathrm{I}}=U-U_{RC}=-\frac{\hbar}{2\pi R C} \ln\left(1+\frac{C}{C_0}\right) \,.
\label{U1}
\end{equation}
In general, for every circuit with a diverging energy we may define a reference circuit by taking the limit $\omega\rightarrow\infty$ where capacitors become short-circuited (and inductors disconnected). The zero-point energy of the reference circuit cannot be calculated with circuit theory in general, but it may be subtracted to get the relative energy. This is analogous to the renormalization of the vdW potential within the dipole approximation, where the diverging Lamb shift of each atom is subtracted \cite{MIL,FoQV}.

When the capacitance $C$ depends on an internal mechanical degree of freedom $\xi$, the potential $U$ leads to a generalized force,
\begin{equation}
f=-\frac{\partial U}{\partial \xi}=-\frac{\partial U}{\partial C}\frac{\partial C}{\partial \xi},
\label{fd}
\end{equation}
acting on $\xi$. Examples include a force between the plates of a parallel-plate capacitor with a plate separation $\xi=y$ (wherein $C\propto1/y$), or a torque in  a rotary variable capacitor with an angle $\xi=\theta$ (Fig.~1c). The generalization to forces in {\it e.g.}\ variable inductors or other quantum piezoelectric devices \cite{CLE,CHU,ROU} should be straightforward.

Unlike the case of vdW/Casimir forces, here a divergence in the potential $U$ implies a similar divergence in the force $f$ (see SM).
Therefore, in the cases where $U$ diverges, we use the renormalized, relative potential inside the expression (\ref{fd}) for the force, leading to a \emph{relative force}. For example, the finite, renormalized result for the relative potential $U_{\mathrm{I}}$ in Eq. (\ref{U1}), leads to a finite result for a relative force $f_{\mathrm{I}}=-\partial U_{\mathrm{I}}/\partial \xi=f-f_{RC}$, which is understood as the difference in the force due to the addition of a capacitor $C_0$ to the reference $RC$ circuit (here $f_{RC}=-\partial U_{RC}/\partial \xi$).  More generally, the relative force is the difference between the force acting inside $C$ due to the full circuit and the force acting inside $C$ due to its reference circuit. When $U$ diverges and has to be renormalized, the relative force is the only force component that can be calculated using macroscopic circuit theory. In the following, we demonstrate the predictive power and generality of such a macroscopic approach.

\subsection*{Examples}

For the resistor and capacitor $C_0$ in series coupled to a parallel-plate capacitor (case I of Fig.~2a), the internal coordinate $\xi\propto 1/C$. With this we obtain from Eq.~(\ref{U1}) a repulsive force $f_{\mathrm{I}}=-\partial U_{\mathrm{I}}/\partial \xi$ (Fig.~2b). Repulsive Casimir forces are normally caused by the phase relationship between the virtual electromagnetic waves reflected inside a Casimir cavity; if the wave is out-of-phase upon reflection on one boundary and in-phase on the other, the force is repulsive. This is the case for electric and magnetic mirrors \cite{Boyer} and was demonstrated \cite{CAPr} for a sandwich of three materials with dielectric constants $\varepsilon_1<\varepsilon_2<\varepsilon_3$, where the Fresnel coefficients change sign. Here we have seen that the electronic environment wherein the capacitor $C$ is embedded may also cause a repulsive force component. Coupling $C$ to another capacitor $C_0$ changes the phases as well (also see SM, Sec. 2). Note however that within our renormalized circuit theory we can only predict the relative force due to $C_0$, which exists on top of the force contributed by the reference $RC$-circuit. In addition, force components not accounted for by circuits theory, such as the standard Casimir force between the capacitor plates, may also matter. It depends on the relative strengths of the force components whether repulsion prevails.

\begin{figure}
\begin{center}
\includegraphics[width=\columnwidth]{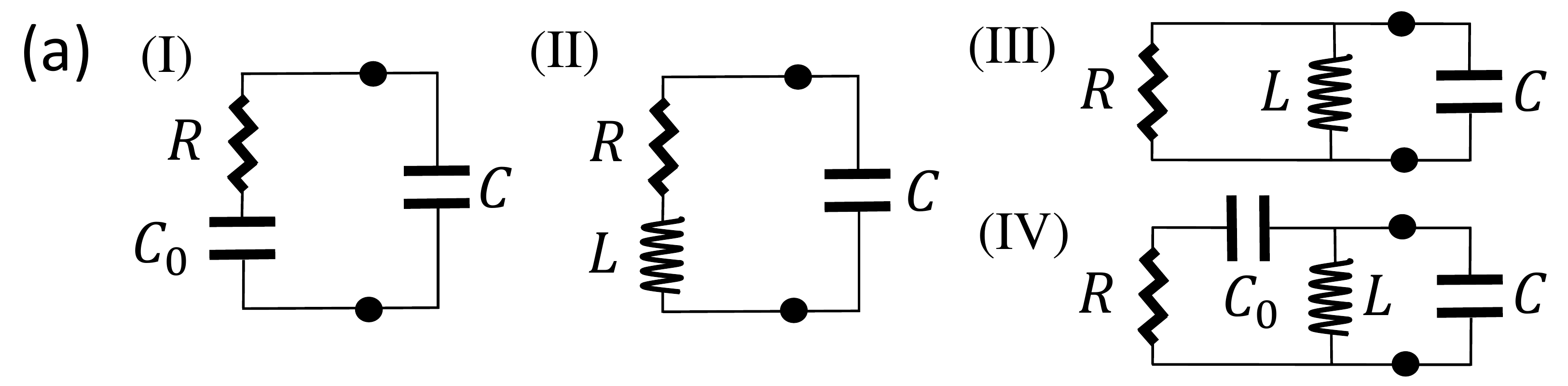}
\includegraphics[width=\columnwidth]{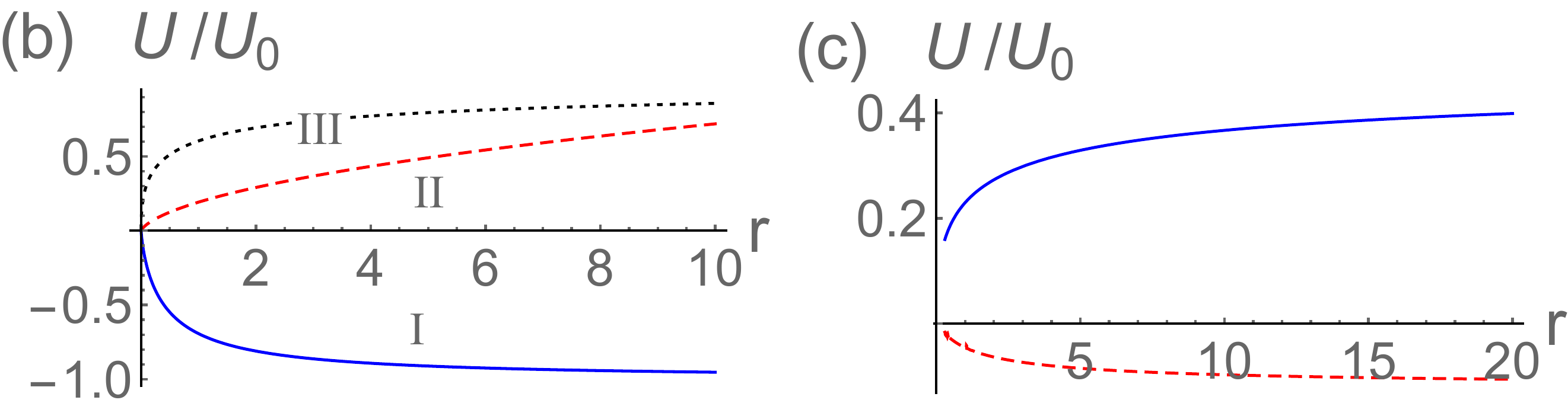}
\caption{\small{
Potential energy inside a capacitor. (a) Four simple cases for the general circuit of Fig.~1b. (b) Potential energy $U$, Eq. (\ref{U}). For case I, $U$ from Eq. (\ref{U1}) is plotted in units of $U_0=\hbar/(2\pi R C_0)$ and as a function of $r=C_0/C$ (solid line). For a parallel-plate capacitor with plate separation $y\propto r$, the resulting relative force between the plates is repulsive. Dashed and dotted lines: same as solid line, for cases II and III with $U_0=\hbar R/L$ and $U_0=(\hbar/2\pi) R/L$, respectively, and $r=L/(C R^2)$, both yielding an attractive force for a plate capacitor. (c) Same as (b) for case IV with $U_0=\hbar/(R C_0)$ and $r=C_0/C$. Depending on the parameter $a=\sqrt{L/C_0}/R$, both attractive ($a=0.5$, solid line) and repulsive ($a=2$, dashed line) potentials are possible.
 }} \label{fig2}
\end{center}
\end{figure}

Consider now an inductor $L$ coupled to the capacitor $C$ (case II in Fig.~\ref{fig2}a). There we have $X=-\omega L$ and obtain from Eq.~(\ref{U}) a finite zero-point energy without the need of renormalization. Convergence is guaranteed by the inductor $L$ in series; its impedance $-i\omega L$ disconnects the circuit at high frequencies. In this case, circuit theory can predict the total Casimir force of the circuit and not only the relative one. As noted above, there may exist an additional contribution to the force beyond the one mediated by the circuit, however this contribution will not depend on the inductor (see below).

Proceeding to the parallel $RLC$ circuit, case III of Fig.~2a, we find the impedance $R(\omega)+iX(\omega)=(\omega L+iR)\times\omega L R/[R^2+(\omega L)^2]$, and renormalize the integral (\ref{U}) by subtracting the $RC$ result. The resulting energy potentials $U_n=\hbar(R/L)F_n$ for both cases $n=\mathrm{II,III}$, are obtained analytically as a function $F_n(r)$ of the parameter $r=L/(C R^2)$, see the Methods section, and are plotted in Fig.~\ref{fig2}b. For $\xi\propto r$, such as in a parallel-plate capacitor, the force now becomes attractive in both cases. At the limits $R\rightarrow 0$ and $R\rightarrow \infty$ for circuits II and III, respectively, the corresponding asymptotic forms, $F_{\mathrm{II}}(r\gg 1), F_{\mathrm{III}}(r\ll 1)\approx\sqrt{r}/4$, both yield the potential, $U\approx (\hbar/4)/\sqrt{LC}$, equal to half the zero-point energy of an isolated $LC$ circuit.

As the last simple example, consider circuit IV of Fig.~\ref{fig2}a. At high frequencies, the inductor is effectively disconnected, approaching circuit I that gave a repulsive force, whereas at very low frequencies $C_0$ is disconnected, leading to an isolated $LC$ circuit with an attractive force. Such competing high and low frequency behaviors both contribute to the potential in Eq. (\ref{U}), which is found as a function of two parameters, $r=C_0/C$ and $a=\sqrt{L/C_0}/R$ (see Methods). Figure 2c presents the potential as a function of $r$ for both $a=0.5$ (solid line) and $a=2$ (dashed line), showing the possibility of attractive and repulsive forces, respectively, for $\xi\propto r$. In such a circuit one can tune the sign of the relative force.

It should be noted that the convergence of the energy of circuit II, due to a series inductor, suggests that the potential (\ref{U}) of any circuit can be made finite without the need of renormalization, by merely considering the finite lengths of wires in the circuit. This is since wires carry an inductance $L_0$ proportional to their length. However, the converging finite result will then depend on $L_0$ and hence on  the specific wire lengths. Moreover, even the arrangement of the wires itself may matter, owing to their finite mutual capacitance. This situation again highlights an essential simplification allowed for by our macroscopic approach: For short enough wire lengths, the predictions for the relative forces are general and independent of such system-specific microscopic details. This idea is further explained and demonstrated in the SM, considering the inductance of finite wires.

\subsection*{Space dependence of forces in a parallel-plate capacitor}

Consider now the specific case of a parallel-plate capacitor, $C=A\varepsilon_0/y$, where the internal coordinate and corresponding force are the separation $\xi=y$ and the force between the plates, respectively ($A$ being the plate area). The plots in Fig. 2b,c as a function of $r\propto 1/C \propto y$ then effectively display the dependence of the potential on the distance $y$, whose derivative is the force. We then observe that the scaling of the potantial/force with the separation $y$ is not universal and crucially depends on the circuit in which the capacitor is embedded.  Not only the sign of the relative force can be altered, but also its functional structure. For example, the asymptotics of the forces (or relative forces) at long distances $y$ are given by
$f_{\mathrm{I}}\propto 1/y^2$, $f_{\mathrm{II}}\propto 1/\sqrt{y}$ and $f_{\mathrm{III}}\propto \ln(y)/y^2$, yielding distinct scalings for the different circuits I, II and III. This behavior is analogous to the modified spatial dependence of the vdW and Casimir forces in different electromagnetic environments: For example, the vdW potential mediated by the one-dimensional photon modes supported by a transmission-line waveguide asymptotically scales as $1/r^3$ whereas the Casimir force scales as $1/r^2$ instead of $1/r^7$ and $1/r^4$, respectively, in free space \cite{QEDFDT,vdWTL} ($r$ being the distance between the interacting objects). In contrast to the standard situation of the Casimir/vdW forces however, where changing the electromagnetic environment amounts to building a totally different and challenging photonic setup (e.g. placing the interacting objects in a waveguide), here this is achieved by varying the impedance of a circuit $Z$ connected to the capacitor. This means that tuning the spatial dependence of zero-point forces in circuits may become as simple as tuning the inductance of a circuit element.

As mentioned above, in addition to the electronic zero-point force, the more standard Casimir force, due to free-space vacuum fluctuations, also exists between the capacitor plates (see SM for more details). It is therefore interesting to compare the two: For the electronic zero-point force we take the series $RLC$ circuit, case II in Fig. 2a, for which we obtained the full force (without renormalization), whereas for the Casimir force we consider the Lifshitz formula, using a plasma model for the metallic capacitor plates \cite{MIL} (SM). Motivated by recent experiments \cite{TEU,LEHN}, we consider aluminium plates at different diameters and plot the resulting forces as a function of plate separation (Fig. 3). Remarkably, the zero-point electronic force can become much stronger at sufficiently large separations, due to its long-range asymptotic scaling, $f_{\mathrm{II}}\propto 1/\sqrt{y}$. Nevertheless, it would be interesting to consider how one may be able to distinguish the electronic zero-point force from the standard Casimir force even in situations where the Casimir force is much stronger. Moreover, as it is well-known from Casimir force measurements, other existing spurious forces, such as those due to electrostatic patch potentials \cite{POD,MUNp,BEHU}, may also mask the electronic force. As discussed below however, these problems turn out to be irrelevant in circuits, where the electronic zero-point force may be measured \emph{independently} of all reference forces.

\subsection*{Measurement as a function of circuit parameters}

One of the most appealing features of electronic circuits, also revealed throughout this work, is their tunability: importantly, this feature may also offer a novel route in the characterization and measurement of zero-point forces. Casimir forces are typically measured and characterized as a function of a distance between interacting objects; or more generally, in our context, as a function of the internal coordinate of the capacitor $\xi$ on which the force acts. However, keeping $\xi$ (and hence $C(\xi)$)  fixed, we may still vary other parameters of the system the capacitor is embedded in, generally represented by its impedance $Z$. Such a possibility typically does not exist in other Casimir-like setups. In practice, this can be achieved by tuning the parameters of circuit components such as variable capacitors, inductors and resistors. While the electronic zero-point force may highly depend on parameters within $Z$, this should not be the case for other forces that act on $\xi$, such as the standard Casimir force induced by the electromagnetic vacuum (SM). Therefore, measuring the force on $\xi$  at a fixed value of $\xi$, but as a function of the circuit parameters in $Z$, may enable distinguishing the electronic zero-point force from all other forces, regardless of their magnitude. This is since these other forces merely contribute a constant offset to the measured force. For the measurement of the relative electronic zero-point force, the same principle applies also for the distinction from the contribution of the reference circuit: e.g. in circuit I, the potential from Eq. (\ref{U1}) is extracted by additionally fixing $R$, such that the contribution of the reference $RC$-circuit to a force measurement as a function of $C_0$ is constant.

The dependence on circuit parameters is illustrated in Fig.~4. For the capacitor $C$, we consider the recently realized superconducting parallel-plate capacitor with a movable plate, cooled to its motional ground state \cite{TEU,LEHN}. The plates, of diameter $15\mu$m, are kept at a fixed separation $\xi=y=50$nm. Forces on the movable plate are balanced by a restoring potential, leading to a small displacement in the separation $y$, which is measured at high precision \cite{TEU}. For circuit I, we plot the repulsive force $f$ between the plates of $C$ as a function of the resistor $R$ and capacitor $C_0$, obtaining $f$ at the fN scale (Fig.~4a). The attractive forces of circuits II and III are plotted as a function of the parameters $R$ and $L$ (Figs.~4b,c). We estimate that these fN-scale forces are detectable with current technology by using the so-called dynamic method \cite{FoQV}. Namely, around each value of a circuit parameter at which the force is to be measured, a weak modulation of this parameter is introduced. This leads to a measurement of the derivative of the force with respect to and as a function of the circuit parameters. Considering the device from Ref. \cite{TEU}, with a displacement sensitivity of $\sim10^{-32}$m$^2$Hz$^{-1}$, we estimate that sub fN-scale forces are detectable, see the Methods section for details. It should also be noted that there is nothing fundamental leading to the fN scale of the forces: this order of magnitude merely reflects the parameters of a specific recently-realized electromechanical setup chosen here to illustrate the predictability and applicability of our approach.

\subsection*{Level shifts in a superconducting qubit}

In addition to forces, the electronic zero-point fluctuations may induce shifts in the quantized energy levels of components such as superconducting qubits (SCQ) embedded in the circuit, in analogy to the Lamb shift in atoms. We consider a SCQ capacitively coupled to an arbitrary circuit represented by a complex impedance $Z$ (Fig.~1d), and study the dependence of the shift on the circuit parameters. We note that a similar effect was observed before in a current-biased Josephson junction connected to an adjustable admittance \cite{DEV1,DEV2}.
For weak coupling $C_g\ll C_J$, with $C_J$ being the total capacitance of the SCQ, the voltage fluctuations at the node $a$ of Fig. 1d, $V_a(t)$, are determined by $Z$ and form an effective reservoir coupled to the charge $Q_J$ of the SCQ, via the Hamiltonian $H_I\approx(C_g/C_J)V_a(t)Q_J$ \cite{DEV,KOCH}. Using lowest order perturbation theory, we find the correction (shift) to the energy difference between the two lowest energy states $|0\rangle$ and $|1\rangle$ of a transmon SCQ (see Methods),
\begin{equation}
\delta\approx \frac{\beta^2}{\hbar Z_J}\int_0^{\infty} \frac{d\omega}{2\pi}S_{V_a}(\omega)\left[\mathrm{P}\frac{\omega_{10}}{\omega_{10}^2-\omega^2}-\frac{1}{\omega+\omega_{21}}\right],
\label{del}
\end{equation}
with the corresponding linewidth $\gamma\approx S_{V_a}(\omega_{10}) \beta^2/(2 \hbar Z_J)$ and the spectrum $S_{V_a}(\omega)=\langle V_a(\omega)V_a(-\omega)\rangle/T_e$.
Here $\hbar\omega_{nm}$ is the energy difference between the levels $n$ and $m$, $\beta=C_g/C_J$, and $Z_J=(\hbar/e^2)\sqrt{E_C/2E_J}$ with $E_C$ and $E_J$ the charging and Josephson energies of the SCQ, and $e$ the electron charge \cite{DEV,KOCH}, while $\mathrm{P}$ denotes Cauchy's principal value.

As a specific example, consider a circuit consisting of a resistor $R$ and a capacitor $C$ in parallel (bottom of Fig. 1d). For weak coupling we find $S_{V_a}(\omega)\approx S_{I_N}(\omega)R^2/(1+\omega^2C^2R^2)$ so that the integral (\ref{del}) converges, yielding $\delta\approx -\omega_0(R/Z_J)\beta^2/(2b)$ for $b\equiv \omega_0 R C\ll 1$, with $\omega_0=\sqrt{8E_C E_J}/\hbar$ (see Methods). We note that in this regime the shift overwhelms the acquired width, $|\delta|/\gamma\approx 1/(2b)\gg 1$. As in the discussion of forces, the shift can be measured as a function of the circuit parameters $R$ and $C$, as illustrated in Fig. 4d, using typical transmon parameters \cite{KOCH}. Shifts of the order of $0.1\%$ of the original qubit resonance, $\omega_{10}=\omega_0-E_C/\hbar$, are realistic.

\section*{Discussion}

Recent studies of quantum phenomena in electrical circuits have indicated their relevance to Casimir physics \cite{WIL,PAR,vdWTL}. Our results suggest that the recent realization of quantum electromechanical systems \cite{TEU,LEHN,CLE,CHU,ROU}, opens more opportunities for the exploration of fluctuation phenomena. A  central theme is the ability to manipulate the electromagnetic environment experienced by a system embedded in a circuit. This results in several important and previously unexplored consequences. First, a variety of tunable fluctuation forces or level shifts may arise, exhibiting for example repulsive or attractive relative forces, with a controllable space dependence. Second, the new possibility to measure these zero-point electronic potentials as a function of the parameters of the environment, rather than as a function of an interaction distance, may entail several advantages: 1. The ability to distinguish these potentials from other random potentials, such as electrostatic patch potentials \cite{POD,MUNp,BEHU}, means that the limitation on measurement precision imposed by the distance dependence of the latter, now becomes irrelevant. 2. The potentially sensitive dynamic measurement method \cite{FoQV} is natural to apply here, where circuit parameters might be simple to modulate.

More generally, this work identifies quantum electromechanical circuits as a promising platform for the exploration of Casimir physics beyond the sphere-plate configuration, which is conceptually different from other recent approaches, {\it e.g.}\ using integrated silicon chips \cite{CHAN} or optical tweezers \cite{NETO}. In this respect, the possibilities allowed for by such systems might go well beyond those found by the above analysis of a few simple circuits, especially considering the generality of our approach. The macroscopic description presented above should allow for the study of novel zero-point forces in a variety of quantum electromechanical devices and circuits, by merely characterising them with an electric impedance, as opposed to more detail-sensitive microscopic approaches \cite{Abinitio}. Considering the potential application of electromechanical devices in quantum hybrid technologies \cite{TEU,CLE,CHU,ROU}, the study of the zero-point potentials therein may also play a role in the characterization and operation of these devices. Moreover, the tunability of the forces found above may prove useful for future nanoelectromechanical systems wherein control of zero-point potentials is essential \cite{CAP4}.

\section*{Methods}

\subsection*{Calculation of zero-point energy and force}
Consider first the circuit (I). The renormalized potential energy is obtained by subtracting from the ``bare" potential $\widetilde{U}_{\mathrm{I}}$, given by Eq. (\ref{U}) with $R(\omega)=R$ and $X(\omega)=1/(\omega C_0)$, the reference $RC$ potential, given by Eq. (\ref{U}) with $R(\omega)=R$ and $X(\omega)=0$, yielding
\begin{equation}
U_{\mathrm{I}}=\widetilde{U}_{\mathrm{I}}-U_{RC}=-\int_{0}^{\infty}\frac{d\omega}{2\pi}\hbar \omega R C    \frac{\frac{C}{C_0}\left(2+\frac{C}{C_0}\right)}     {\left(1+\frac{C}{C_0}\right)^2+(RC\omega)^2\left[1+\left(1+\frac{C}{C_0}\right)^2\right]+(R C\omega)^4}.
\label{integral}
\end{equation}
Upon calculating the integral analytically we arrive at Eq.~(\ref{U1}). The integral (\ref{integral}) begins converging for $\omega\gg 1/(RC)$, where it behaves as $\int d\omega /\omega^3$. Since the lumped-element circuit theory we use is essentially a low-frequency theory valid for $\omega\ll c/l$, the convergence is meaningful only if it begins for $\omega$-values smaller than the cutoff of the theory ($l$ being a typical size of a circuit element and $c$ the speed of light) \cite{FoQV}. This leads to the validity condition, $1/(RC)\ll c/l$, which is satisfied for the parameters in Fig. 4a (using e.g. $l= 10\mu$m).

We calculate analytically the potentials in circuits (II) and (III) in a similar fashion, recalling that circuit II does not require renormalization, and obtain
\begin{eqnarray}
U_{\mathrm{II}}&=&\hbar \frac{R}{L}\frac{q}{4\sqrt{4q-1}}\left[1+\frac{2}{\pi} \arctan \left(\frac{2q-1}{\sqrt{4q-1}}\right)\right],
\nonumber\\
U_{\mathrm{III}}&=&\hbar\frac{R}{L}\frac{q}{8\pi}\left[\frac{\left(\frac{4}{q}-2\right)}{\sqrt{\frac{4}{q}-1}}\arctan\left(\frac{\frac{2}{q}-1}{\sqrt{\frac{4}{q}-1}}\right)+\pi\left(\frac{2}{q}-1\right)\sqrt{\frac{1}{\frac{4}{q}-1}}+2\ln q\right],
\label{U23}
\end{eqnarray}
with the dimensionless circuit parameter $q=L/(R^2 C)$. The convergence properties of the integrals that lead to these results, in analogy to those of the integral (\ref{integral}) for circuit I, give the validity conditions $c/l \gg1/\sqrt{LC},R/L$ (circuit II) and $c/l\gg 1/\sqrt{LC},1/(RC)$ (circuit III), which are again satisfied for the parameters in Figs. 4b and 4c.
The above analytical results were also verified by comparing them to a direct numerical evaluation of the integral as a function of $q$.
For circuit (IV) we obtain the renormalized potential in a converging integral form, and as a function of the dimensionless parameters $r=C_0/C$ and $a=\sqrt{L/C_0}/R$,
\begin{equation}
U_{\mathrm{IV}}=\frac{\hbar}{R C_0}\frac{r}{4\pi}\int_0^{\infty} dx \frac{x\left\{2a^2r(r+1)-r^2-x a^2r[2(a^2-1)r^2+a^2r]\right\}-1}{(1+x)\left\{r^2 x(1-a^2 r x)^2+[1-x a^2 r(r+1)]^2\right\}}.
\label{U4b}
\end{equation}
This integral can be evaluated numerically for different values of $r$ and $a$. Its convergence properties imply the validity condition $c/l\gg 1/\sqrt{LC},1/(RC)$.

In order to calculate the force in Fig.~4, we consider a parallel-plate capacitor with a plate area $A$ and separation $y$, where $C=A \varepsilon/y$.
The force is then found either analytically by a simple differentiation, or numerically by differentiating the potential in its integral form and performing the resulting integral numerically.

\subsection*{Measurement of forces in the dynamical method}
Considering the capacitor plate realization from Ref. \cite{TEU}, the forces from Fig. 4, that are calculated using the physical parameters of this plate-capacitor realization, act on the drumhead mechanical mode of the capacitor. This mechanical mode is modeled as an harmonic oscillator with amplitude $x(t)$, resonant frequency $\Omega$ and damping rate $\Gamma$.
The zero-point electronic force $f(Z)$ acting on the plate depends on a parameter of the circuit, generally denoted by $Z$. For a given circuit with a constant $Z$, the resulting static force will shift the plate by
$x_{dc}(Z)=f(Z)/(m \Omega^2)$, where $m$ is the mass of the mechanical oscillator. Therefore, in principle, one can  infer the function $f(Z)$ by measuring the plate position $x$ as a function of $Z$, provided that the signal value $x_{dc}(Z)$ is larger than the standard deviation of the noise on $x$. This is the essence of the static method. In the dynamic method \cite{FoQV}, we further consider a weak modulation of the circuit parameter  around its value $Z$ in which we wish to measure, $\delta Z(t)=u\cos(\omega_p t)$, with $u\ll Z$ and $\omega_p$ being the modulation frequency. Solving for $x(t)$ influenced by the force $f[Z+\delta Z(t)]\approx f(Z)+\delta Z(t)\left.\frac{\partial f}{\partial Z}\right|_Z$, we find the displacement signal for a modulation with a bandwidth $B\ll \Gamma$ around the mechanical resonance $\omega_p=\Omega$,
\begin{equation}
x_s \equiv \left[\int_{\omega_p-\frac{B}{2}}^{\omega_p+\frac{B}{2}}\frac{d\omega}{2\pi}\left|x(\omega)\right|^2\right]^{\frac{1}{2}}=\sqrt{\pi} Q \eta(Z) x_{dc}(Z).
\label{xs}
\end{equation}
Here $Q=\Omega/\Gamma$ is the quality factor of the mechanical resonance and $\eta(Z)=u \left.\frac{\partial f}{\partial Z}\right|_Z/f(Z)\ll 1$. In order to be able to measure the force, this signal has to be much larger than the noise in $x$ contained in the bandwidth $B$.

Considering the $f\sim1$ fN scale forces from Fig.~4 and the physical parameters of the electromechanical capacitor, $m=48$ pg, $\Omega=2\pi\times10.56$ MHz and $Q=3.3\times10^5$ \cite{TEU}, we find static displacements of the order $x_{dc}\approx4.7\times10^{-18}$m, much smaller than the mechanical zero-point motion of the plate, $x_{zp}=4.1\times10^{-15}$m, making the static method irrelevant.
Turning to the dynamical method, the displacement sensitivity (noise spectrum) of the device in Ref. \cite{TEU}, which depends on the magnitude of a probe signal used in the measurement, reaches an optimum of  $5.5\times10^{-34}$m$^2$Hz$^{-1}$. Considering non-optimal operation, we take $S_{N}(\omega)\sim 10^{-32}$m$^2$Hz$^{-1}$ over the bandwidth of interest. The noise signal, $x_N=\sqrt{S_N B}$, for a bandwidth of {\it e.g.} $B=0.01\Gamma$, is $x_N=1.4\times10^{-16}$m. Comparing it to the signal, Eq. (\ref{xs}), with a modulation strength of {\it e.g.} $\eta=0.05$, we find $x_s\approx1.4\times10^{-13}$m, three orders of magnitude larger than the noise. Such a signal-to-noise ratio is likely to be sufficient in order to resolve forces like those shown in Fig.~4.

\subsection*{Calculation of level shifts in a superconducting qubit}

Following the Hamiltonian description of circuits \cite{DEV}, the relevant degrees of freedom of the circuit in Fig. 1d are the nodes $a$ and $J$, representing the impedance $Z$ and the superconducting qubit (SCQ), respectively.  Each circuit node is represented by a dynamical variable (operator), {\it e.g.}\ the voltage $\hat{V}_m(t)$ ($m=a,J$), or more commonly the ``flux'' defined by $\hat{\phi}_m(t)=\int_0^t dt' \hat{V}_m(t')$. Then, the interaction between the systems $Z$ and $J$, coupled by the capacitor $C_g$, is given by \cite{DEV,KOCH}
\begin{equation}
H_{ZJ}=\beta\hat{V}_a \hat{Q}_J,
\end{equation}
where $\hat{Q}_J$ is the canonical conjugate of $\hat{\phi}_J$, $\beta=C_g/(C_g+C_J)\approx C_g/C_J$ and $C_J\gg C_g$ is the total capacitance of the SCQ. Moving to the interaction picture, the Hamiltonian is $H_I(t)=\beta \widetilde{V}_a(t) \widetilde{Q}_J(t)$, where the tilded operators are dynamically evolved using the individual free dynamics (Hamiltonian) of each system. For the system $Z$ this means that $\widetilde{V}_a(t)$ may in fact be found by solving the Kirchhoff equations for the voltage $V_a(\omega)=\int dte^{i\omega t}\widetilde{V}_a(t)$ in the limit $C_g\rightarrow 0$, obtaining the spectrum  $S_{V_a}(\omega)=\langle V_a(\omega)V_a(-\omega)\rangle/T_e$. Then, for weak coupling $\beta\ll 1$, we treat the circuit/impedance $Z$ as a reservoir, and obtain the lowest-order non-Hermitian correction it induces on the SCQ Hamiltonian (see Supplementary Materials),
\begin{equation}
H_{\mathrm{eff}}=\sum_{n,m}E_{nm}^{(2)}|n\rangle\langle m|, \quad E_{nm}^{(2)}=\frac{\beta^2}{\hbar}\sum_{n'}Q_{nn'}Q_{n'm}\left[-\Delta(\omega_{nn'})-\frac{i}{2}S_{V_a}(\omega_{nn'})\right].
\label{Heff}
\end{equation}
Here the pairs of states $n,m$ are either identical or degenerate, $Q_{nn'}=\langle n|\hat{Q}_J|n'\rangle$, and $\Delta(\omega)=\frac{1}{2\pi}\mathrm{P}\int_{-\infty}^{\infty}d\omega'\frac{S_{V_a}(\omega')}{\omega'-\omega}$. For a transmon SCQ we have \cite{KOCH}
\begin{equation}
|Q_{nn'}|\approx2e\left(\frac{E_J}{8E_C}\right)^{\frac{1}{4}}\left[ \delta_{n',n+1}\sqrt{\frac{n+1}{2}}+\delta_{n',n-1}\sqrt{\frac{n}{2}}\right],
\label{tr}
\end{equation}
and $E_{n+1}-E_n=\sqrt{8E_CE_J}-E_C(n+1)$, yielding the shift, $\delta=\mathrm{Re}\{E^{(2)}_{11}-E^{(2)}_{00}\}$, and the width correction, $\gamma=-2\mathrm{Im}\{E^{(2)}_{11}\}$, from Eq. (\ref{del}) in the main text. Finally, considering the specific case of the $RC$ circuit (bottom of Fig.~1d), we easily find $V_a(\omega)=I_N(\omega) R/(1-i\omega C R)$, so that $S_{V_a}(\omega>0)=2\hbar\omega R/(1+\omega^2C^2R^2)$, leading to the results from the main text (see Supplementary Materials for more details).

\begin{figure}
\begin{center}
  \includegraphics[width=\columnwidth]{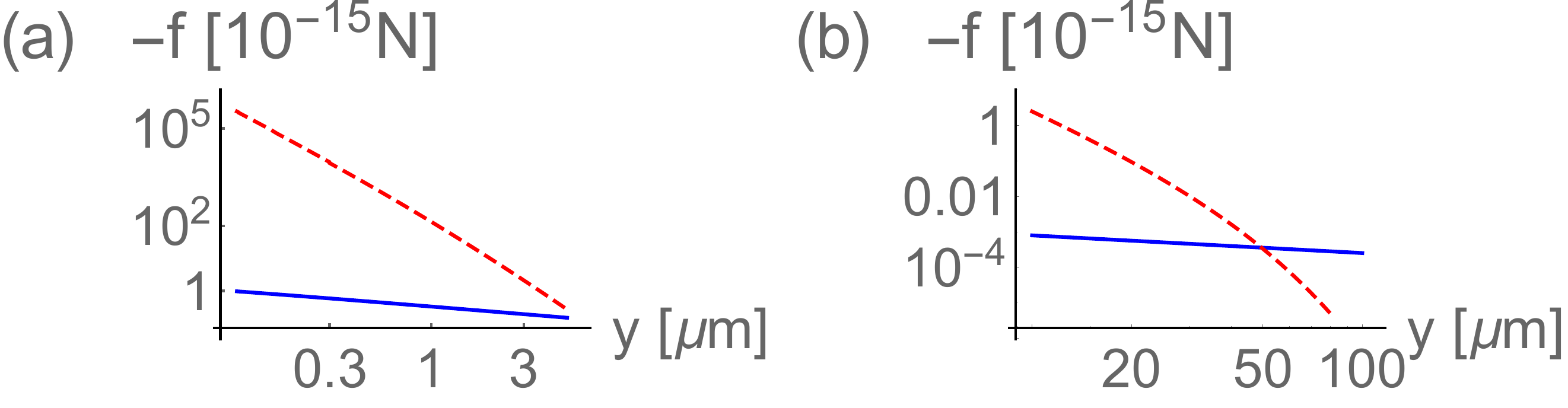}
\caption{\small{
Comparison between the electronic zero-point force (blue solid line) and the standard Casimir force (red dashed line) acting between the plates of a parallel-plate capacitor. The capacitor with a capacitance $C=A\varepsilon_0/y$ is assumed to be connected in series to a resistor, $R=10\mathrm{\Omega}$, and an inductor, $L=0.1$nH (circuit II of Fig. 2a). The forces are plotted as a function of the plate separation $y$, and the plate diameter is taken to be: (a) $15\mu$m, and (b) $200\mu$m. Due to its long-range scaling, the electronic zero-point force can become much stronger [see case (b) and text].
 }} \label{fig3}
\end{center}
\end{figure}

\begin{figure}
\begin{center}
  \includegraphics[width=\columnwidth]{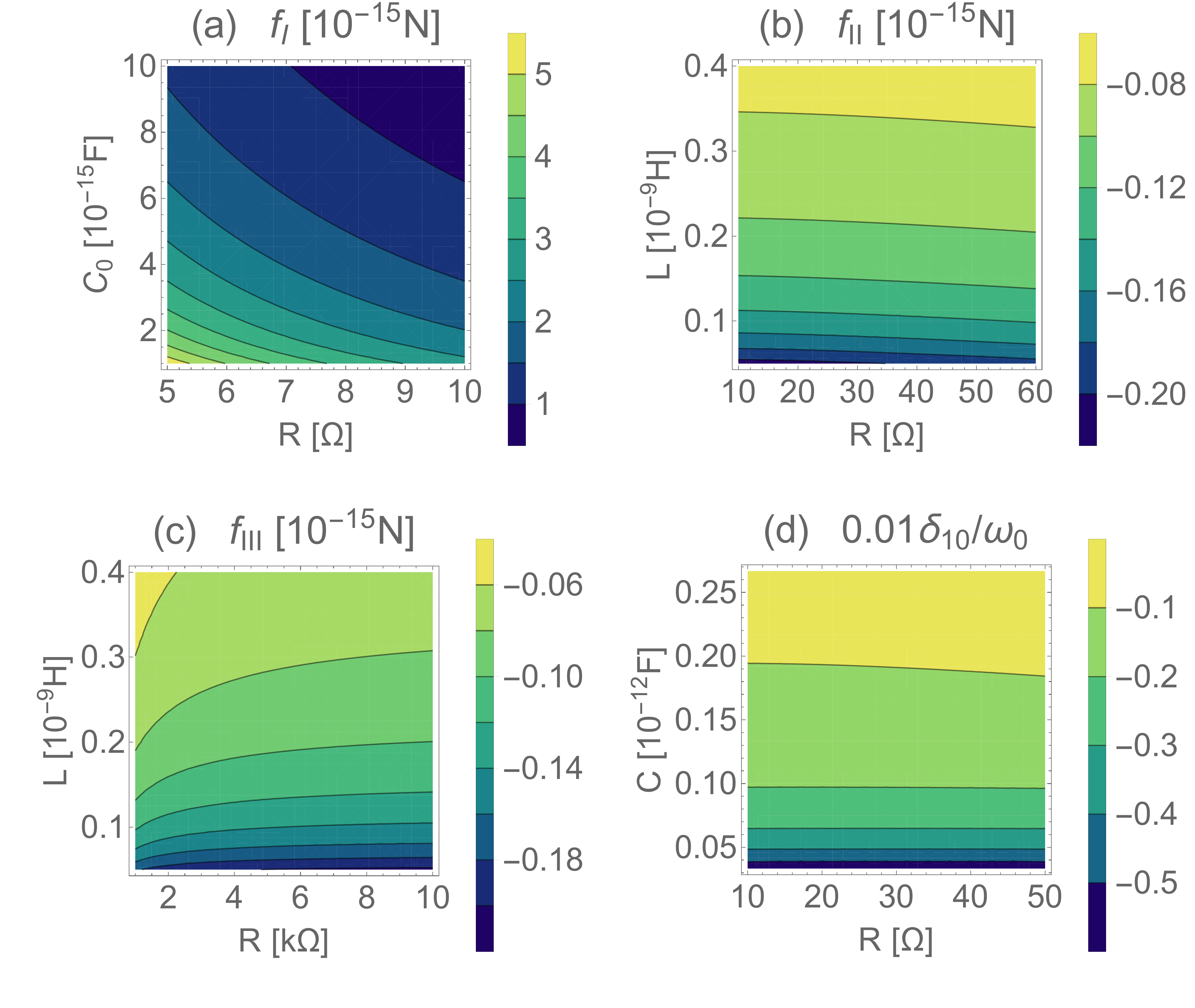}
\caption{\small{
The ability to measure fluctuation potentials as a function of parameters of the environment (circuit) presents an interesting possibility opened by circuits for fluctuation-induced phenomena. (a,b,c): The force between the capacitor plates in the circuits (I,II,III) from Fig. 2a is plotted as a function of circuit parameters. Circuit (I) exhibits repulsive relative forces whereas circuits (II) and (III) exhibit attractive forces. In the limits of small $R$ for $f_{\mathrm{II}}$ and large $R$ for $f_{\mathrm{III}}$, both forces become identical to that in the isolated $LC$ circuit. All three cases are plotted using the physical parameters of the parallel-plate electromechanical capacitor of Ref. \cite{TEU}, whose displacement-measurement sensitivity of $\sim 10^{-32}$m$^2$Hz$^{-1}$ is estimated to be sufficient to detect the sub fN forces we find (text).
(d) Shift in the transition frequency of a superconducting transmon qubit with $\sqrt{E_C/8E_J}=0.1$, $C_g/C_J=0.1$ and $\omega_0=2\pi\times5$GHz. Shifts of the order of $0.1\%$ of the original transition frequency are observed, much larger than the acquired level width.
 }} \label{fig4}
\end{center}
\end{figure}

\end{document}